\overfullrule=0pt
\input harvmac
\input epsf.sty

\def\a{{\alpha}}

\def\l{{\lambda}}

\def\b{{\beta}}

\def\g{{\gamma}}

\def\d{{\delta}}
\def\e{{\epsilon}}

\def\half{{1\over 2}}
\def\p{{\partial}}

\def\t{{\theta}}

\def\({\left(}
\def\){\right)}

\def\cF{{\cal F}}


 \Title{\vbox{\hbox{AEI-2010-124}}} {\vbox{
     \centerline{\bf Towards Field Theory Amplitudes From the Cohomology }
     \centerline{\bf of Pure Spinor Superspace }
  }
 }

\bigskip\centerline{Carlos R. Mafra\foot{email: crmafra@aei.mpg.de}}

\bigskip

\centerline{\it }
\medskip
\centerline{\it Max-Planck-Institut f\"ur
Gravitationsphysik, Albert-Einstein-Institut}
\smallskip
\centerline{\it 14476 Golm, Germany}
\medskip

\vskip .3in

A simple BRST-closed expression for the color-ordered super-Yang-Mills 5-point 
amplitude at tree-level is proposed in pure spinor superspace and
shown to be BRST-equivalent to the field theory limit of the
open superstring 5-pt amplitude. It is manifestly cyclic invariant and each one
of its five terms can be associated to the 
five Feynman diagrams 
which use only cubic vertices. Its form also suggests an empirical method
to find superspace expressions in the cohomology of the pure spinor BRST operator
for higher-point amplitudes based on their kinematic pole structure. Using this
method, Ans\"atze for the 6- and 7-point 10D super-Yang-Mills
amplitudes which map to their 14 and 42 color-ordered diagrams are conjectured
and their 6- and 7-gluon expansions are explicitly computed.

\vskip .3in

\Date {July 2010}

\lref\mafraids{
  C.~R.~Mafra,
  ``Pure Spinor Superspace Identities for Massless Four-point Kinematic
  Factors,''
  JHEP {\bf 0804}, 093 (2008)
  [arXiv:0801.0580 [hep-th]].
}

\lref\mafrabcj{
  C.~R.~Mafra,
  ``Simplifying the Tree-level Superstring Massless Five-point Amplitude,''
  JHEP {\bf 1001}, 007 (2010)
  [arXiv:0909.5206 [hep-th]].
}

\lref\bcj{
  Z.~Bern, J.~J.~M.~Carrasco and H.~Johansson,
  ``New Relations for Gauge-Theory Amplitudes,''
  Phys.\ Rev.\  D {\bf 78}, 085011 (2008)
  [arXiv:0805.3993 [hep-ph]].
}

\lref\tye{
  H.~Tye and Y.~Zhang,
  ``Dual Identities inside the Gluon and the Graviton Scattering Amplitudes,''
  arXiv:1003.1732 [hep-th].
}
\lref\vanhove{
  N.~E.~J.~Bjerrum-Bohr, P.~H.~Damgaard, T.~Sondergaard and P.~Vanhove,
  ``Monodromy and Jacobi-like Relations for Color-Ordered Amplitudes,''
  arXiv:1003.2403 [hep-th].
}
\lref\Medinas{
  R.~Medina, F.~T.~Brandt and F.~R.~Machado,
  ``The open superstring 5-point amplitude revisited,''
  JHEP {\bf 0207}, 071 (2002)
  [arXiv:hep-th/0208121].
\semi
  L.~A.~Barreiro and R.~Medina,
  ``5-field terms in the open superstring effective action,''
  JHEP {\bf 0503}, 055 (2005)
  [arXiv:hep-th/0503182].
}
\lref\mafraids{
  C.~R.~Mafra,
  ``Pure Spinor Superspace Identities for Massless Four-point Kinematic
  Factors,''
  JHEP {\bf 0804}, 093 (2008)
  [arXiv:0801.0580 [hep-th]].
}
\lref\tese{
  C.~R.~Mafra,
  ``Superstring Scattering Amplitudes with the Pure Spinor Formalism,''
  arXiv:0902.1552 [hep-th].
}
\lref\ictp{
  N.~Berkovits,
  ``ICTP lectures on covariant quantization of the superstring,''
  arXiv:hep-th/0209059.
}
\lref\stie{
  S.~Stieberger and T.~R.~Taylor,
  ``Multi-gluon scattering in open superstring theory,''
  Phys.\ Rev.\  D {\bf 74}, 126007 (2006)
  [arXiv:hep-th/0609175].
}
\lref\stieS{
  S.~Stieberger and T.~R.~Taylor,
  ``Supersymmetry Relations and MHV Amplitudes in Superstring Theory,''
  Nucl.\ Phys.\  B {\bf 793}, 83 (2008)
  [arXiv:0708.0574 [hep-th]].
}
\lref\siegel{
  W.~Siegel,
  ``Classical Superstring Mechanics,''
  Nucl.\ Phys.\  B {\bf 263}, 93 (1986).
}
\lref\mhv{
  S.~J.~Parke and T.~R.~Taylor,
  ``An Amplitude for $n$ Gluon Scattering,''
  Phys.\ Rev.\ Lett.\  {\bf 56}, 2459 (1986).
}
\lref\PSsuperspace{
  N.~Berkovits,
  ``Explaining pure spinor superspace,''
  arXiv:hep-th/0612021.
}
\lref\hennTree{
  J.~M.~Drummond and J.~M.~Henn,
  ``All tree-level amplitudes in N=4 SYM,''
  JHEP {\bf 0904}, 018 (2009)
  [arXiv:0808.2475 [hep-th]].
}
\lref\witten{
  E.~Witten,
  ``Twistor - Like Transform In Ten-Dimensions,''
  Nucl.\ Phys.\  B {\bf 266}, 245 (1986).
}
\lref\psf{
  N.~Berkovits,
  ``Super-Poincare covariant quantization of the superstring,''
  JHEP {\bf 0004}, 018 (2000)
  [arXiv:hep-th/0001035].
}
\lref\cheung{
  C.~Cheung, D.~O'Connell and B.~Wecht,
  ``BCFW Recursion Relations and String Theory,''
  arXiv:1002.4674 [hep-th].
}
\lref\boels{
  R.~H.~Boels, D.~Marmiroli and N.~A.~Obers,
  ``On-shell Recursion in String Theory,''
  arXiv:1002.5029 [hep-th].
}
\lref\bigHowe{
  P.~S.~Howe,
  ``Pure Spinors Lines In Superspace And Ten-Dimensional Supersymmetric
  Theories,''
  Phys.\ Lett.\  B {\bf 258}, 141 (1991)
  [Addendum-ibid.\  B {\bf 259}, 511 (1991)].
\semi
  P.~S.~Howe,
  ``Pure Spinors, Function Superspaces And Supergravity Theories In
  Ten-Dimensions And Eleven-Dimensions,''
  Phys.\ Lett.\  B {\bf 273}, 90 (1991).
}
\lref\multiloop{
  N.~Berkovits,
  ``Multiloop amplitudes and vanishing theorems using the pure spinor
  formalism for the superstring,''
  JHEP {\bf 0409}, 047 (2004)
  [arXiv:hep-th/0406055].
}
\lref\anomaly{
  N.~Berkovits and C.~R.~Mafra,
  ``Some superstring amplitude computations with the non-minimal pure spinor
  formalism,''
  JHEP {\bf 0611}, 079 (2006)
  [arXiv:hep-th/0607187].
}
\lref\Rq{
  G.~Policastro and D.~Tsimpis,
  ``R**4, purified,''
  Class.\ Quant.\ Grav.\  {\bf 23}, 4753 (2006)
  [arXiv:hep-th/0603165].
}
\lref\FORM{
  J.~A.~M.~Vermaseren,
  ``New features of FORM,''
  arXiv:math-ph/0010025.
\semi
  M.~Tentyukov and J.~A.~M.~Vermaseren,
  ``The multithreaded version of FORM,''
  arXiv:hep-ph/0702279.
}
\lref\PSS{
  C.~R.~Mafra,
  ``PSS: A FORM Program to Evaluate Pure Spinor Superspace Expressions,''
  arXiv:1007.4999 [hep-th].
}
\lref\treelevel{
  N.~Berkovits and B.~C.~Vallilo,
  ``Consistency of super-Poincare covariant superstring tree amplitudes,''
  JHEP {\bf 0007}, 015 (2000)
  [arXiv:hep-th/0004171].
}
\lref\CSW{
  F.~Cachazo, P.~Svrcek and E.~Witten,
  ``MHV vertices and tree amplitudes in gauge theory,''
  JHEP {\bf 0409}, 006 (2004)
  [arXiv:hep-th/0403047].
}
\lref\ampStie{
  D.~Oprisa and S.~Stieberger,
  ``Six gluon open superstring disk amplitude, multiple hypergeometric  series
  and Euler-Zagier sums,''
  arXiv:hep-th/0509042.
}

\lref\WIP{C.~Mafra, O.~Schlotterer, S.~Stieberger, D.~Tsimpis,
work in progress}
\lref\tsimpis{
	G.~Policastro and D.~Tsimpis,
        ``R**4, purified,''
	Class.\ Quant.\ Grav.\  {\bf 23}, 4753 (2006)
	[arXiv:hep-th/0603165].
}
\lref\humbertotwo{
  H.~Gomez and C.~R.~Mafra,
  ``The Overall Coefficient of the Two-loop Superstring Amplitude Using Pure
  Spinors,''
  JHEP {\bf 1005}, 017 (2010)
  [arXiv:1003.0678 [hep-th]].
}
\newsec{Introduction}

As Parke and Taylor have shown for MHV amplitudes \mhv, it is sometimes possible
to obtain simple expressions for seemingly complicated Yang-Mills amplitudes in four spacetime
dimensions. Using the pure spinor formalism \psf\ and its pure spinor superspace 
\PSsuperspace\ (see also \witten)
it will be proved that the tree-level color-ordered
five-point super-Yang-Mills amplitude
in ten dimensions can be written
simply as
\eqn\Fivepts{
{\cal A}_5(1,2,3,4,5) = {\langle L_{45}L_{12}V^3\rangle \over s_{45}s_{12}} + {\rm cyclic}(12345),
}
where $V^j$ is the unintegrated massless vertex operator and $L_{ij}$ is related to the OPE
of a unintegrated and an integrated vertex operator in a way to be defined below.

It will also be suggested that higher-point amplitudes might have simple forms like the
above, as there seems to be a direct correspondence between superspace expressions
and Feynman diagrams which use only cubic vertices as in the arguments of \bcj. Using the
empirical method described in subsection 3.1,
it will be argued that the super-Yang-Mills 6- and 7-point color-ordered amplitudes are
proportional to
\eqn\sixptAnsatzI{
{\cal A}_6(1,2,3,4,5,6) = 
{\langle L_{12}L_{34}L_{56}\rangle\over 3 s_1s_3s_5}
}
$$
+\half {\langle T_{123}\over s_1 t_1}{(V^4L_{56}\over s_5} + {L_{45}V^6)\rangle \over s_4} 
-\half {\langle T_{126}\over s_1 t_3}{(V^3L_{45}\over s_4} + {L_{34}V^5)\rangle \over s_3}
	     + {\rm cyclic}(1{\ldots} 6)
$$
and 
\eqn\sevenptsI{
{\cal A}_7(1,2,3,4,5,6,7) =
 {\langle T_{231} L_{45}L_{67} \rangle \over s_2 t_1 s_4 s_6}
+ {\langle T_{123} T_{564} V_7  \rangle \over s_1 t_1 s_5 t_4}
+ {\langle T_{127} T_{345} V_6  \rangle \over s_1 t_7 s_3 t_3}
}
$$
- {\langle T_{123}T_{456}V_7 \rangle \over s_1 t_1 s_4t_4}
- {\langle T_{127}T_{453}V_6 \rangle\over s_1 t_7 s_4t_3}
- {\langle T_{123}L_{45} L_{67} \rangle \over s_1 t_1 s_4 s_6}
+ {\rm cyclic(1{\ldots} 7)}
$$
where $T_{ijk}$ is related to the OPE of one unintegrated and two integrated vertices
in a way to be defined below and $s_1,{\ldots} ,s_6$ and $t_1,{\ldots} ,t_3$ ($s_1,{\ldots} ,s_7$ and $t_1,{\ldots} ,t_7$)
are the 6-point (7-point) generalized Mandelstam variables of \refs{\stie,\stieS}.
Using a computer program \PSS, the 6- and 7-gluon
expansions 
of \sixptAnsatzI\ and \sevenptsI\
are computed in Appendix B\foot{In the amplitude
computations of \refs{\stie,\stieS} the results were written in the 4D helicity
formalism language, so a 10D comparison of results is not straightforward. However
a comparison to the result \ampStie\ should be made \WIP. After the first version
of this paper came out, the 6-gluon
amplitude has been successfully matched against the results of Zvi Bern, which
he kindly provided \ref\zvi{Zvi Bern, private communication.}. The 7-gluon amplitude still remains to be checked.}.

Furthermore, given
that the tree-level SYM 4-point amplitude can be written as \mafrabcj
\eqn\Fourpts{
{\cal A}_4(1,2,3,4) = {1\over s_{12}}\langle L_{12} V^3V^4 \rangle + {1\over s_{41}}\langle L_{41} V^2V^3 \rangle,
}
it is pointed out that the four-point Jacobi-like Bern-Carrasco-Johansson kinematic identity \bcj\ becomes
\eqn\fourptbcj{
\langle L_{\{12}V_{3\}} V_4\rangle = 0,
}
where $\{ijk\}$ means a sum over cyclic permutations of $(ijk)$. Its vanishing is
explained by noting that it is BRST trivial.
For the five-point amplitude \Fivepts, the generalized BCJ identities of \refs{\tye,\vanhove}
hold in the form of
\eqn\ebcju{
-{L_{45}\over s_{45}}L_{\{12}V_{3\}} 
+{L_{42}\over s_{24}}L_{\{13}V_{5\}}
-{L_{12}\over s_{12}}L_{\{34}V_{5\}}
+{L_{51}\over s_{51}}L_{\{23}V_{4\}} = 0,
}
etc. It is well-known that there are powerful four-dimensional methods to compute scattering amplitudes 
recursively (see \hennTree\ and references therein).
The hints of a simplified ten-dimensional
parametrization of field theory tree-level amplitudes using pure spinors\foot{It was suggested a long
time ago that pure spinors simplify the description of super-Yang-Mills and supergravity theories \bigHowe.
The superspace results obtained with the pure spinor formalism seem to realize those expectations.}
seem to suggest that there might be similar methods in a ten-dimensional pure spinor superspace 
setup -- which is desirable since there is no need to differentiate between MHV and NMHV contributions
as in the four-dimensional methods.

This paper is organized as follows. In section 2 an ansatz will be given for the tree-level
five-point SYM amplitude by analogy with the structure of the known four-point amplitude. In section
3 the five-point ansatz will be derived from the field theory limit of a BRST-equivalent 
expression of the superstring amplitude computed in \mafrabcj. In subsection 3.1 an empirical method
to write down similar Ans\"atze for higher-point amplitudes is presented, and expressions for the
6- and 7-point super-Yang-Mills amplitudes in ten-dimensional space-time are conjectured. In Appendix A
the BCJ kinematic relations and its generalization \refs{\tye,\vanhove} are written down using the pure spinor
representations of the previous sections. Finally, in Appendix B the first few terms of the
(rather long) 5-, 6- and 7-gluon expansions from \Fivepts, \sixptAnsatzI\ and \sevenptsI\
are written down
(the full expansions can be easily generated with a computer using \PSS\ or other methods).

\newsec{Tree-level amplitudes with the pure spinor formalism}

The prescription to compute n-point tree-level open string 
amplitudes with the pure spinor formalism is given 
by \psf\foot{For background material
in the pure spinor formalism, see \refs{\ictp,\tese}. 
The conventions for the OPE's however follow the appendix A of \mafrabcj.}
\eqn\treepresc{
{\cal A}_n = \langle V^1(0)V^{(n-1)}(1)V^n(\infty) \int dz_2 U^2(z_2) 
{\ldots} \int dz_{(n-2)}U^{(n-2)}(z_{(n-2)}\rangle,
}
where $V^i(z) = \l^\a A^i_\a $ and 
$U^i(z) = \p\t^\a A^i_\a + \Pi^m A^i_m + d_\a W^\a_i + \half {\cal F}_{mn}^i N^{mn}$
are the unintegrated and integrated vertices with conformal weight zero and one, respectively,
and $i$ is the label denoting the different strings being scattered.
The massless sector of the open superstring is described by
the ten-dimensionl super-Yang-Mills superfields $[A_\a,A_m,W^\a,{\cal F}_{mn}]$ which satisfy
the equations of motion \refs{\witten,\multiloop,\tese},
\eqn\SYMEOM{
Q{\cal F}_{mn} = 2k_{[m} (\l\g_{n]} W), \;
Q W^{\a} = {1\over 4}(\l\g^{mn})^{\a}{\cal F}_{mn},\;
QA_m = (\l\g_m W) + k_m(\l A),\; Q V = 0,
}
where $\l^\a(z)$ is a pure spinor satisfying $\l^\a \g^m_{\a\b} \l^\b = 0$,
$Q=\l^\a D_\a$ is the pure spinor BRST operator and $D_\a = \p_\a + \half k_m (\g^m \t)_\a$
is the supersymmetric derivative\foot{In what follows spinor index contractions are denoted
by parenthesis, e.g. $\l^\a D_\a = (\l D)$ and the worldsheet positions
are mostly omitted.}. They have the following $\theta$-expansions,
\ref\thetaSYM{
  	J.~P.~Harnad and S.~Shnider,
	``Constraints And Field Equations For Ten-Dimensional Superyang-Mills
  	Theory,''
  	Commun.\ Math.\ Phys.\  {\bf 106}, 183 (1986).
\semi
	H.~Ooguri, J.~Rahmfeld, H.~Robins and J.~Tannenhauser,
        ``Holography in superspace,''
        JHEP {\bf 0007}, 045 (2000)
        [arXiv:hep-th/0007104].
\semi
	P.~A.~Grassi and L.~Tamassia,
        ``Vertex operators for closed superstrings,''
        JHEP {\bf 0407}, 071 (2004)
        [arXiv:hep-th/0405072].
}\tsimpis\
$$
A_{\a}(x,\t)={1\over 2}a_m(\g^m\t)_\a -{1\over 3}(\xi\g_m\t)(\g^m\t)_\a
-{1\over 32}F_{mn}(\g_p\t)_\a (\t\g^{mnp}\t) + \ldots
$$
$$
A_{m}(x,\t) = a_m - (\xi\g_m\t) - {1\over 8}(\t\g_m\g^{pq}\t)F_{pq}
         + {1\over 12}(\t\g_m\g^{pq}\t)(\p_p\xi\g_q\t) + \ldots
$$
$$
W^{\a}(x,\t) = \xi^{\a} - {1\over 4}(\g^{mn}\t)^{\a} F_{mn}
           + {1\over 4}(\g^{mn}\t)^{\a}(\p_m\xi\g_n\t)
	   + {1\over 48}(\g^{mn}\t)^{\a}(\t\g_n\g^{pq}\t)\p_m F_{pq} 
	   + \ldots
$$
\eqn\expansions{
\cF_{mn}(x,\t) = F_{mn} - 2(\p_{[m}\xi\g_{n]}\t) + {1\over
4}(\t\g_{[m}\g^{pq}\t)\p_{n]}F_{pq} + {\ldots},
}
where $a_m(x) = e_m {\rm e}^{ik\cdot x}$, $\xi^{\a}(x) =\chi^\a {\rm e}^{ik\cdot x}$ are the bosonic
and fermionic polarizations and $F_{mn} = 2\p_{[m} a_{n]}$ is the field-strength.

After using the OPE's to eliminate the conformal weight-one variables from \treepresc, the
integration of the zero-modes of $\l^\a$ and $\t^\a$ is carried out by taking only
the terms which contain three $\l$'s and five $\t$'s in the correlator which are proportional
to the pure spinor measure
\eqn\tlct{
\langle (\l\g^m\t)(\l\g^n\t)(\l\g^p\t)(\t\g_{mnp}\t)\rangle = 1,
}
where the normalization can be chosen arbitrarily\foot{See however the tree-level, one-loop and
two-loop calculations of \humbertotwo\ to check how the choice has to be taken into account at
higher-loops.}.
The normalization condition \tlct\ defines the action of
the pure spinor angle-brackets $\langle \phantom{n}\rangle$. Arbitrary 
{\it pure spinor superspace} expressions are written down as
\eqn\pssexpr{
\langle \l^\a\l^\b \l^\g f_{\a\b\g}(\t)\rangle,
}
where $f_{\a\b\g}(\t)$ is given in terms of super-Yang-Mills superfields, e.g. $f_{\a\b\g}(\t) =
A^i_\a(\t) A^j_\b(\t) A^k_\g(\t)$. The measure \tlct\ is in the cohomology of the pure spinor
BRST operator and can not be written as the supersymmetry variation of a BRST-closed object, so
amplitudes computed from \treepresc\ are supersymmetric \psf.

As an illustration of the above steps, the supersymmetric tree-level 3-point amplitude following from
\treepresc\ is given by\foot{One also has to evaluate the functional integration of the
exponentials $\prod :{\rm e}^{ik^i\cdot X(z_i)}:$, but they will not appear explicitly in this paper.}
\eqn\three{
{\cal A}_3 =\langle (\l A^1)(\l A^2)(\l A^3)\rangle.
}
Evaluating the explicit component expansion for {\it e.g.} the 3-gluon amplitude, is a matter of
plugging in the expansions \expansions\ and selecting the components with five $\t$'s which
contain the gluon fields. Doing that one obtains,
\eqn\threegluon{
{\cal A}_3 =
- {1\over 64}\(
k^3_m e^1_{r}e^2_{s} e^3_{n} -
k^2_m e^1_{r}e^2_{n} e^3_{s} +
k^1_m e^1_{n}e^2_{r} e^3_{s}
\)
\langle (\l \g^r\t) (\l \g^s\t)(\l \g_p \t)(\t\gamma^{pmn}\t)\rangle.
}
As mentioned in the appendix of \anomaly, symmetry arguments and the
normalization condition \tlct\ fix all pure spinor correlators.
Among the list of \anomaly\ one finds
$$
\langle (\l \g^r\t) (\l \g^s\t)(\l \g_p \t)(\t\gamma^{pmn}\t)\rangle 
= {1\over 120}\d^{rsp}_{pmn} = {1\over 45}\d^{rs}_{mn},
$$
so the 3-gluon amplitude \threegluon\ is given by
\eqn\tgfim{
{\cal A}_3 = - {1\over 2880}\( 
  (e^1\cdot e^2)(k^2 \cdot e^3) 
+ (e^1\cdot e^3)(k^1 \cdot e^2)
+ (e^2\cdot e^3)(k^3 \cdot e^1)\).
}
Given the systematic nature of the above procedure,
an implementation using {\tt FORM} \FORM\ has been written which performs
these expansions automatically \PSS. So although component
expansions can have many thousand terms as in the 7-gluon amplitude
discussed in appendix B, they come from much simpler superspace 
expressions which can be analysed by hand.

\newsec{The 5-pt field theory amplitude ansatz}

When the amplitude involves more than three strings, the prescription
\treepresc\ requires the computation of the OPE's with integrated 
vertices. In this section we will be concerned with the field theory
limit (FT) of the string scattering. The 5-point FT amplitude 
will be given an Ansatz motivated by the superspace form of the FT
4-point amplitude, which will later be obtained from a BRST equivalent
expression of the first principles superstring 5-point amplitude 
evaluated in \mafrabcj.

In superspace, the
OPE between the unintegrated and integrated vertex operators 
is given by 
$V^i(z)U^j(w) \rightarrow {{\tilde L}_{ij}\over z-w}$, with \mafraids
\eqn\Lij{
{\tilde L}_{ij}(\t) = A^i_m(\l\g^m W^j) + (\l A^i)(k^i\cdot A^j).
}
Using the equations of motion \SYMEOM\ it follows that 
\eqn\trivial{
Q{\tilde L}_{ij} = -s_{ij} (\l A^i)(\l A^j), \quad Q(A^i\cdot A^j) = {\tilde L}_{ij} + {\tilde L}_{ji} \equiv 2 {\tilde L}_{(ij)}
}
where\foot{Note that the usual definition for massless particles is $s_{ij} = 2(k^i\cdot k^j)$.}
$s_{ij} = (k^i\cdot k^j)$.
Using \trivial\ 
and defining $L_{ij} = 1/2({\tilde L}_{ij} - {\tilde L}_{ji})$
the superfield ${\tilde L}_{ij}$ can be written as\foot{I thank Dimitrios Tsimpis 
for suggesting the separation of the BRST-trivial
part of ${\tilde L}_{ij}$.}
\eqn\LA{
{\tilde L}_{ij} = L_{ij} + {1\over 2}Q(A^i\cdot A^j).
}

The massless 4-point super-Yang-Mills amplitude
obtained from the field theory limit of the open string amplitude is given by \mafrabcj\
\eqn\quapt{
{\cal A}(1,2,3,4) = {1\over s_{12}}\langle {\tilde L}_{12} V^3V^4 \rangle + {1\over s_{41}}\langle {\tilde L}_{41} V^2V^3 \rangle
= {1\over s_{12}}\langle  L_{12} V^3V^4 \rangle + {1\over s_{41}}\langle L_{41} V^2V^3 \rangle
}
where we used that $\langle Q(A^i\cdot A^j)V^k V^l\rangle = 0$, which follows from
integrating the BRST charge by parts. 
The other sub-amplitudes are obtained from \quapt\ by relabeling,
$$
{\cal A}(1,3,4,2) = - {1\over s_{13}}\langle L_{13} V^2V^4 \rangle - {1\over s_{12}}\langle L_{12} V^3V^4 \rangle
$$
\eqn\quaptfim{
{\cal A}(1,4,2,3) = - {1\over s_{14}}\langle L_{41} V^2V^3 \rangle + {1\over s_{13}}\langle L_{13} V^2V^4 \rangle.
}
It is easy to check that the amplitudes in \quaptfim\ are BRST-closed.

As emphasized in \bcj, a color-ordered 5-point tree-level amplitude consists of five
diagrams with purely cubic vertices specifying the poles,
\eqn\bcjF{
{\cal A}(1,2,3,4,5) = 
  {n_1 \over s_{45}s_{12}}
+ {n_2 \over s_{51}s_{23}}
+ {n_3 \over s_{12}s_{34}}
+ {n_4 \over s_{23}s_{45}}
+ {n_5 \over s_{34}s_{51}}.
}
As the BRST variation of $L_{ij}$ is proportional to $s_{ij}$, the idea now is to
construct a pure spinor superspace expression using
$L_{ij}$ and $L_{kl}$ in the numerators of the terms containing poles in $s_{ij}$ and
$s_{kl}$, in such a way as to obtain a BRST-closed expression. 
It is straightforward to see that the amplitudes
$$
{\cal A}(1,2,3,4,5) = 
  {\langle L_{45}L_{12} V^3 \rangle \over s_{45}s_{12}}
+ {\langle L_{51}L_{23} V^4 \rangle \over s_{51}s_{23}}
+ {\langle L_{12}L_{34} V^5 \rangle \over s_{12}s_{34}}
+ {\langle L_{23}L_{45} V^1 \rangle \over s_{23}s_{45}}
+ {\langle L_{34}L_{51} V^2 \rangle \over s_{34}s_{51}}
$$
$$
{\cal A}(1,3,2,4,5) = 
  {\langle L_{45}L_{13} V^2 \rangle \over s_{45}s_{13}}
- {\langle L_{51}L_{23} V^4 \rangle \over s_{51}s_{23}}
- {\langle L_{13}L_{42} V^5 \rangle \over s_{13}s_{24}}
- {\langle L_{23}L_{45} V^1 \rangle \over s_{23}s_{45}}
- {\langle L_{42}L_{51} V^3 \rangle \over s_{24}s_{51}}
$$
$$
{\cal A}(1,4,3,2,5) = 
  {\langle L_{25}L_{14} V^3 \rangle \over s_{25}s_{14}}
+ {\langle L_{34}L_{51} V^2 \rangle \over s_{51}s_{43}}
+ {\langle L_{23}L_{14} V^5 \rangle \over s_{14}s_{32}}
+ {\langle L_{25}L_{34} V^1 \rangle \over s_{43}s_{25}}
+ {\langle L_{51}L_{23} V^4 \rangle \over s_{32}s_{51}}
$$
$$
{\cal A}(1,3,4,2,5) = 
  {\langle L_{25}L_{13} V^4 \rangle \over s_{25}s_{13}}
- {\langle L_{34}L_{51} V^2 \rangle \over s_{51}s_{34}}
+ {\langle L_{13}L_{42} V^5 \rangle \over s_{13}s_{42}}
- {\langle L_{25}L_{34} V^1 \rangle \over s_{34}s_{25}}
+ {\langle L_{42}L_{51} V^3 \rangle \over s_{42}s_{51}}
$$
$$
{\cal A}(1,2,4,3,5) = 
  {\langle L_{35}L_{12} V^4 \rangle \over s_{35}s_{12}}
+ {\langle L_{42}L_{51} V^3 \rangle \over s_{51}s_{43}}
- {\langle L_{12}L_{34} V^5 \rangle \over s_{12}s_{43}}
+ {\langle L_{35}L_{42} V^1 \rangle \over s_{42}s_{35}}
- {\langle L_{34}L_{51} V^2 \rangle \over s_{43}s_{51}}
$$
\eqn\ansatz{
{\cal A}(1,4,2,3,5) =
  {\langle L_{35}L_{14} V^2 \rangle \over s_{35}s_{14}}
- {\langle L_{42}L_{51} V^3 \rangle \over s_{51}s_{24}}
- {\langle L_{23}L_{14} V^5 \rangle \over s_{14}s_{23}}
- {\langle L_{35}L_{42} V^1 \rangle \over s_{24}s_{35}}
- {\langle L_{51}L_{23} V^4 \rangle \over s_{23}s_{51}}
}
are BRST-closed. One can also check that all sub-amplitudes in \ansatz\ 
are related to $A(1,2,3,4,5)$ by index relabeling, taking into account the 
antisymmetry of $L_{ij}$ and its fermionic nature. The signs in \ansatz\ precisely
match the ones presented in equation (4.5) of \bcj, so one can identify
$$
n_1 = \langle L_{45}L_{12}V^3\rangle,\;
n_2 = \langle L_{51}L_{23}V^4\rangle,\;
n_3 = \langle L_{12}L_{34}V^5\rangle,\;
n_4 = \langle L_{23}L_{45}V^1\rangle
$$
$$
n_5 = \langle L_{34}L_{51}V^2\rangle,\;
n_6 = \langle L_{25}L_{14}V^3\rangle,\;
n_7 = \langle L_{23}L_{14}V^5\rangle,\;
n_8 = \langle L_{25}L_{34}V^1\rangle
$$
$$
n_9 =    \langle L_{25}L_{13}V^4\rangle,\;
n_{10} = \langle L_{13}L_{42}V^5\rangle,\;
n_{11} = \langle L_{42}L_{51}V^3\rangle,\;
n_{12} = \langle L_{35}L_{12}V^4\rangle
$$
\eqn\mapping{
n_{13} = \langle L_{35}L_{42}V^1\rangle,\quad
n_{14} = \langle L_{35}L_{14}V^2\rangle,\quad
n_{15} = \langle L_{45}L_{13}V^2\rangle.
}
As will be mentioned in the appendix, the above ``solution'' for the $n_i$'s of \bcj\ 
do not satisfy the strict Bern-Carrasco-Johansson (BCJ) kinematic identities, but they do satisfy
the generalized BCJ's of \refs{\tye,\vanhove}. As explained in \refs{\tye,\vanhove}, a general
parametrization of the sub-amplitudes in terms of poles does not necessarily satisfy the BCJ Jacobi-like
identities of \bcj. They must however satisfy ``generalized BCJ identities'', for which
the original BCJ relations are just one out of many possible solutions.

The amplitudes in \ansatz\ will now be obtained
from the field theory limit of
a BRST-equivalent expression of the pure spinor superstring amplitude computed in \mafrabcj.

\newsec{First principles derivation of the 5-pt ansatz \ansatz}

The massless 5-point open superstring amplitude is given by \mafrabcj\foot{The notation here slightly
differs from \mafrabcj, but should not lead to confusion.}
$$
{\cal A}_5(1,2,3,4,5) =   \langle L_{2131}V^4V^5 \rangle K_1 - \langle L_{2134}V^5\rangle K_2 
- \langle L_{2434}V^1V^5\rangle K'_1 + \langle L_{2431}V^5 \rangle K_3
$$
\eqn\ampcpt{
- \langle L_{2331}V^4V^5 \rangle K_5 - \langle L_{2334}V^1V^5 \rangle K'_4 + \langle D_{23}V^1V^4V^5 \rangle (1+s_{23})K_6,
}
where $K_j$ and $K'_j$ denote integrals which satisfy \Medinas
$$
s_{34}K_2 = s_{13}K_1 + s_{23}K_4, \quad s_{24}K_3 = s_{12}K_1 - s_{23}K_5, \quad K_1 = K_4 - K_5
$$
$$
s_{12}K_2 = s_{24}K'_1 + s_{23}K'_4, \quad s_{13}K_3 = s_{34}K'_1 - s_{23}K'_5,  \quad K'_1 = K'_4 - K'_5
$$
\eqn\relat{
(1+s_{23})K_6 = s_{34}K'_4 - s_{13}K_5
= s_{12}K_4 - s_{24}K'_5.
}
The various $L_{ijkl}$ kinematic building blocks have the following
expressions\foot{In the computations of \mafrabcj\ there were terms 
with factors of $(A^i W^j)V^k$ in the expressions for $L_{jiki}$.
But it was shown that using the relations \relat\
those terms drop out from the amplitude, so they are not
written in this paper for brevity.}
\eqn\Ldutu{
L_{2131} =
+  {\tilde L}_{12}((k^1+k^2)\cdot A^3)
+  (\l\g^{m}W^3) \big[  A^1_m(k^1\cdot A^2) +  A^{1\,n}{\cal F}^2_{mn}
       - (W^1\g_m W^2)\big]
}
\eqn\kinematics{
L_{2134} = {\tilde L}_{12}{\tilde L}_{43}, \quad
D_{23} = - (A^2\cdot A^3).
}
Relabeling $1\leftrightarrow 4$ determines $L_{2434}$ from \Ldutu\ and $L_{2431}$ from
\kinematics. Finally, the OPE identities of \mafrabcj\ (which are related to the BCJ dualities
of \bcj) imply that
\eqn\SBCJ{
L_{2331} = L_{3121} - L_{2131}, \quad
L_{2334} = L_{3424} - L_{2434},
}
which are used to obtain the remaining kinematic factors appearing 
in \ampcpt\ from the expression for \Ldutu\ and relabelings thereof.


Using the integral relation for $K_6$ and the expression for $D_{23}$,
$$
\langle D_{23}V^1V^4V^5 \rangle K_6 = -(1+s_{23})K_6\langle (A^2\cdot A^3)V^1V^4V^5\rangle 
= (s_{13}K_5 -s_{34}K'_4)\langle (A^2\cdot A^3)V^1V^4V^5\rangle
$$
the amplitude \ampcpt\ becomes
$$
{\cal A}_5(1,2,3,4,5) =   \langle L_{2131}V^4V^5\rangle K_1 - \langle L_{2134} V^5\rangle K_2 
- \langle L_{2434}V^1V^5\rangle K'_1 + \langle L_{2431}V^5\rangle K_3
$$
\eqn\manip{
- \langle (L_{2331} - s_{13}(A^2\cdot A^3)V^1) V^4V^5 \rangle K_5 
- \langle(L_{2334} - s_{34}(A^2\cdot A^3)V^4) V^1V^5\rangle K'_4.
}
A key point is to note from \Ldutu\ is that it obeys the identity
\eqn\key{
QL_{2131} = s_{12}\({\tilde L}_{23} V_1  - {\tilde L}_{13}V_1  + {\tilde L}_{12} V_3 \)
- (s_{12}+s_{13}+s_{23}){\tilde L}_{12}V_3,
}
and by defining\foot{I thank Dimitrios Tsimpis for suggesting the relevance of using this definition
in the context of an ansatz for the 6-pt amplitude. It turns out to clean up the 5-pt formul{\ae} too.}
\eqn\Tdef{
T_{ijk} \equiv L_{jiki} - S_{jiki}, \quad
S_{jiki} = {1\over 2}s_{ij}( (A^j\cdot A^k)V^i  - (A^i\cdot A^k)V^j)
- {1\over 2}(s_{ik}+s_{jk})(A^i\cdot A^j)V^k,
}
the BRST-trivial parts from $L_{jiki}$ are removed and one obtains a BRST variation
written in terms of $L_{ij}$ instead of ${\tilde L}_{ij}$,
\eqn\QT{
QT_{ijk} = s_{ij} L_{\{ij}V_{k\}} - (s_{jk}+s_{ki}+s_{ij}) L_{ij}V_k.
}
Furthermore, using \QT\ it is easy to show that $Q(T_{jik} - T_{jki} - T_{kij}) = 0$. In
fact this combination is BRST-trivial,
\eqn\BRSTtrivial{
T_{jik} - T_{jki} - T_{kij} = Q\( (A^i\cdot A^j)(k^i\cdot A^k) - (A^i\cdot A^k)(k^i\cdot A^j)
- (A^j\cdot A^k)(k^k\cdot A^i)\).
}
Using the definitions \LA, \Tdef, the relations \relat\ obeyed by the integrals and
the identity \BRSTtrivial\ the superstring five point amplitude \manip\ becomes
\eqn\manipt{
{\cal A}_5(1,2,3,4,5) = 
   \langle L_{12}L_{34}V_5\rangle K_2
 + \langle L_{13}L_{24}V_5\rangle K_3
}
$$
 + \langle T_{123} V_4V_5\rangle K_1 
 - \langle T_{432} V_1V_5\rangle K'_1
 + \langle T_{321} V_4V_5\rangle K_5
 - \langle T_{234} V_1V_5\rangle K'_5.
$$
As discussed in \Medinas, under the twist
$2\leftrightarrow 3$ and $1\leftrightarrow 4$ of the vertex operators on the disc,
the integrals behave as
\eqn\twist{
K_1 \leftrightarrow K'_1, \quad K_4 \leftrightarrow K'_4, \quad K_5 \leftrightarrow K'_5, \quad
K_2 \leftrightarrow K_2, \quad K_3 \leftrightarrow K_3,
}
from which one can easily check that the 5-pt superstring amplitude \manipt\ is anti-symmetric,
as it should on general grounds.

Writing the five point integrals in the two dimensional basis $(T,K_3)$ of \Medinas\ where
\eqn\Tint{
T = s_{12}s_{34}K_2 + (s_{12}s_{51} - s_{12}s_{34} + s_{34}s_{45})K_3
}
as follows \mafrabcj\
\eqn\Ku{
K_1 = {T\over s_{12} s_{45} } - \({s_{34} \over s_{12}} + {s_{23} \over s_{45}}\)K_3, \quad
K'_1 = {T\over s_{34} s_{51} } - \({s_{12} \over s_{34}} + {s_{23} \over s_{51}}\)K_3
}
\eqn\Kc{
K_5 = {T\over s_{23} s_{45} } - \({s_{12} \over s_{45}} + {s_{51} \over s_{23}} -1 \)K_3, \quad
K'_5 = {T\over s_{23} s_{51} } - \({s_{34} \over s_{51}} + {s_{45} \over s_{23}} -1 \)K_3
}
the amplitude \manipt\ becomes
\eqn\amp{
{\cal A}_5(1,2,3,4,5) = T\, A_{\rm YM}(\t) + K_3\, A_{F^4}(\t),
}
where,
\eqn\AYMT{      
A_{\rm YM}(\t)=                      
           {\langle T_{123} V^4 V^5\rangle  \over s_{12} s_{45}}
          - {\langle T_{234} V^1 V^5\rangle \over s_{23} s_{51}}
	  + {\langle L_{12} L_{34} V^5\rangle \over s_{12} s_{34}}
          + {\langle T_{321} V^4 V^5\rangle \over s_{23} s_{45}}
          - {\langle T_{432} V^1 V^5\rangle \over s_{34} s_{51}}
}
and
\eqn\AFqT{
A_{F^4}(\t) = \langle L_{12} L_{34} V^5\rangle
	  + \langle L_{13} L_{24} V^5\rangle
	  - \langle T_{234} V^1 V^5\rangle
	  + \langle T_{321} V^4 V^5\rangle
}
$$
          - \langle L_{12} L_{34} V^5\rangle \( {s_{45}\over s_{12}} + {s_{51}\over s_{34}}\)
          - \langle T_{123} V^4 V^5  \rangle \( {s_{34}\over s_{12}} + {s_{23}\over s_{45}}\)
          + \langle T_{234} V^1 V^5  \rangle \( {s_{45}\over s_{23}} + {s_{34}\over s_{51}}\)
$$
$$
          - \langle T_{321} V^4 V^5 \rangle \( {s_{51}\over s_{23}} + {s_{12}\over s_{45}}\)	  
          + \langle T_{432} V^1 V^5 \rangle \( {s_{23}\over s_{51}} + {s_{12}\over s_{34}}\).
$$
One can also find
a BRST-equivalent form for the amplitude by using the fact that
$Q(L_{mn}/s_{mn}) = - V^mV^n$ to rewrite $\langle T_{ijk}V^m V^n\rangle$ as
$-\langle T_{ijk} Q(L_{mn}/s_{mn})\rangle$, which upon integration of the
BRST charge by parts using \QT\ implies that
\eqn\implies{
\langle T_{ijk} V_m V_m \rangle = - \langle {L_{mn}\over s_{mn}} (s_{ij}L_{\{ij}V_{k\}} - s_{ijk}L_{ij}V_k)\rangle.
}
A somewhat tedious but straightforward use of \implies\ in the expressions \AYMT\ and \AFqT\ allows them
to be rewritten as
\eqn\AYM{
A_{\rm YM}(\t) = 
           {\langle L_{45}L_{12}V^3 \rangle\over s_{45}s_{12}}
	  + {\langle L_{51}L_{23}V^4 \rangle\over s_{51}s_{23}}
          + {\langle L_{12}L_{34}V^5 \rangle\over s_{12} s_{34}}
	  + {\langle L_{23}L_{45}V^1 \rangle\over s_{23}s_{45}}
	  + {\langle L_{34}L_{51}V^2 \rangle\over s_{34}s_{51}}	  
}
and
\eqn\AFq{
A_{F^4}(\t) = 
- \langle L_{45}L_{12}V^3 \rangle\( {s_{23} \over s_{45}} + {s_{34} \over s_{12}}\)
- \langle L_{51}L_{23}V^4 \rangle\( {s_{34} \over s_{15}} + {s_{45} \over s_{23}}\)
}
$$          	  
	  - \langle L_{12}L_{34}V^5 \rangle\( {s_{45} \over s_{12}} + {s_{51} \over s_{34}}\)
	  - \langle L_{23}L_{45}V^1 \rangle\( {s_{51} \over s_{23}} + {s_{12} \over s_{45}}\)
	  - \langle L_{34}L_{51}V^2 \rangle\( {s_{12} \over s_{34}} + {s_{23} \over s_{51}}\)
$$
$$
+ \langle L_{12}L_{34}V^5 + L_{51}L_{23}V^4
              - L_{13}L_{42}V^5 + L_{23}L_{45}V^1\rangle
+{s_{13}\over s_{51}}\langle L_{51}L_{\{23}V_{4\}}\rangle
-{s_{24}\over s_{45}}\langle L_{45}L_{\{12}V_{3\}}\rangle.
$$
In the field theory limit $T\rightarrow 1$ and $K_3 \rightarrow 0$ \Medinas, so the
first principles derivation of \ansatz\ is completed. The 5-gluon component expansion
were already computed in \mafrabcj, and shown to agree with earlier RNS results \Medinas.

\subsec{Higher-point amplitudes}

It is worth checking whether the simple mappings between the cubic Feynman diagrams
and pure spinor building blocks persist at higher-points.
The discussion in section 2 suggests a way to write down
n-point field theory amplitudes.
For each one of the $2^{n-2}(2n-5)!!/(n-1)!$ color-ordered
diagrams specifying the kinematic poles \bcj, a ghost-number-three
numerator whose BRST
transformation is proportional to those poles should be written down. One then tries
to find a combination with the correct dimension of a n-point amplitude 
such that the sum of all diagrams is BRST-closed.

To help finding candidates for superfield building blocks, the first principles
tree-level superstring amplitude prescription \refs{\psf,\treelevel}
can be used as guide. For example, 
the superfield ${\tilde L}_{ij}$
appears in the OPE of $V^i(z)U^j(w)$ in the 4-pt string amplitude \mafraids,
and its BRST transformation $Q{\tilde L}_{ij} = -s_{ij}V^iV^j$ 
has precisely the Mandelstam variable to cancel poles in the 5-pt amplitude.
Similarly, the superfield $L_{jiki}$ comes from the numerator of the $1/z_{ij}z_{ik}$ pole
in the OPE $V^i(z_i) U^j(z_j) U^k(z_k)$ appearing in the 5-pt computation \mafrabcj, and
its BRST transformation has the required Mandelstam variables to cancel
poles in the 6-pt amplitude,
\eqn\QL{
QL_{jiki} = s_{ij}({\tilde L}_{jk}V^i -{\tilde L}_{ik}V^j + {\tilde L}_{ij}V^k)
- (s_{jk}+s_{ki}+s_{ij}){\tilde L}_{ij}V^k.
}
As the expressions must be in the cohomology of the pure spinor BRST operator,
one also removes the BRST-trivial parts of the building blocks ${\tilde L}_{ij}$ and 
$L_{jiki}$, using $L_{ij}$ and $T_{ijk}$ instead. 

Following the above procedure for the 14 color-ordered diagrams of
the 6-point amplitude which are generated from the cyclic permutations of
the diagrams in Figures 1, 2 and 3, a BRST-closed expression with the correct pole structure
looks like\foot{I thank Oliver 
Schlotterer and Dimitrios Tsimpis for many valuable discussions.}
\eqn\sixptAnsatz{
{\cal A}_6(1,2,3,4,5,6) = 
{\langle L_{12}L_{34}L_{56}\rangle\over 3 s_1s_3s_5}
}
$$
+\half {\langle T_{123}\over s_1 t_1}{(V^4L_{56}\over s_5} + {L_{45}V^6)\rangle \over s_4}
-\half {\langle T_{126}\over s_1 t_3}{(V^3L_{45}\over s_4} + {L_{34}V^5)\rangle \over s_3}
	     + {\rm cyclic}(1{\ldots} 6)
$$
where $s_1 = s_{12}, s_2 = s_{23}, {\ldots} ,s_6 = s_{61}$, $t_1 = (s_{12}+s_{23}+s_{13})$,
$t_2 = (s_{23}+s_{34}+s_{24})$ and $t_3 = (s_{34}+s_{45}+s_{35})$ are the 6-point
Mandelstam variables of \stie. The full component expansion for the 6-gluon amplitude obtained from
\sixptAnsatz\ contains 6706 terms \PSS\ and it was checked to be gauge invariant\foot{After the first version
of this paper appeared, Zvi Bern kindly provided his Mathematica file with the field theory 6-gluon amplitude
written in terms of polarization and momenta.
A perfect match was obtained.}. The first few
terms of this expansion are given in Appendix B.

\smallskip
\centerline{\epsfxsize 1.4truein\epsfbox{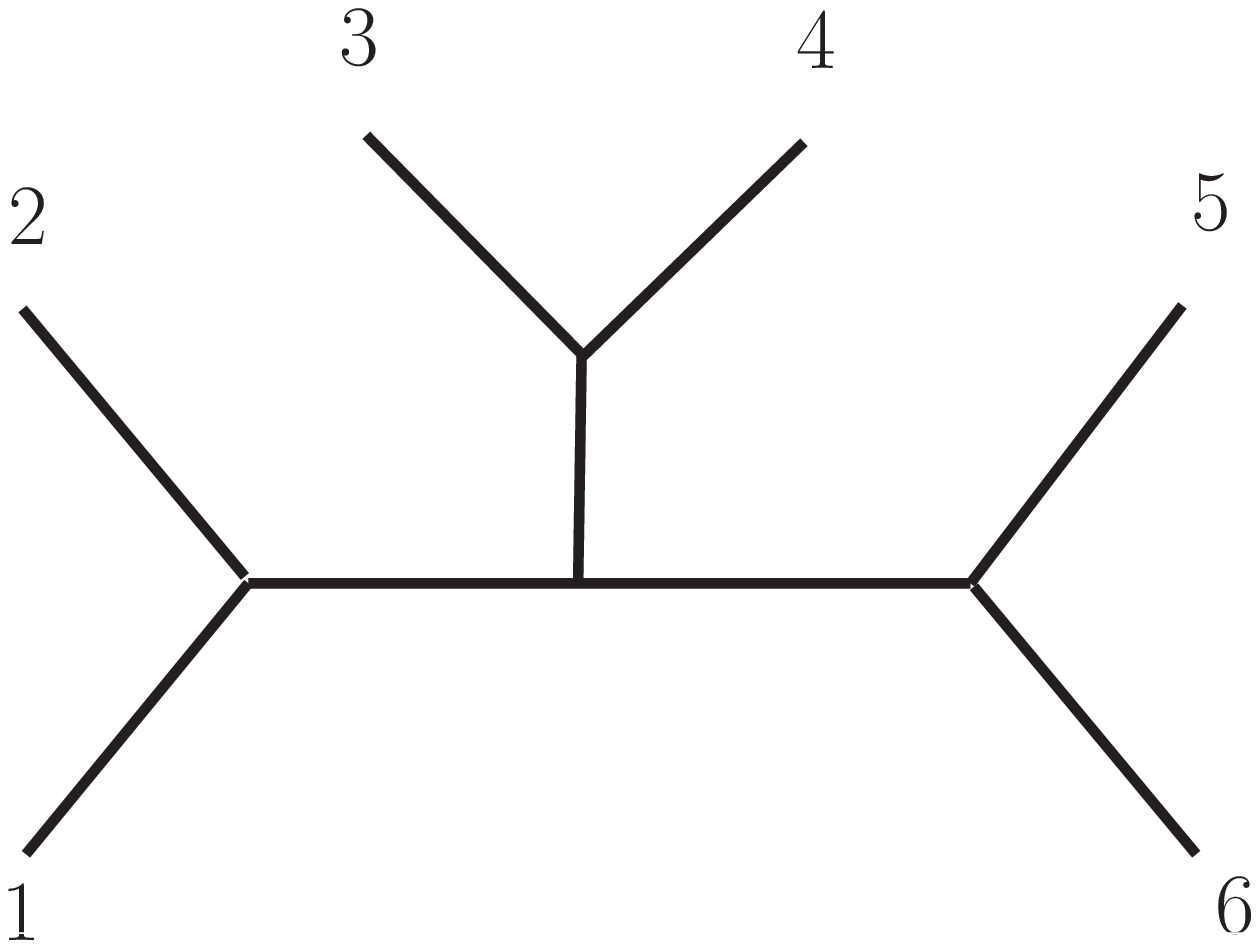}}
\centerline{\ninepoint \baselineskip=2pt {\bf Fig. 1.} {The diagram associated 
with $\langle {L_{12}\over s_1}{L_{34}\over s_3}{L_{56}\over s_5}\rangle$.}}
\smallskip
\bigskip
\centerline{\epsfxsize 3.4truein\epsfbox{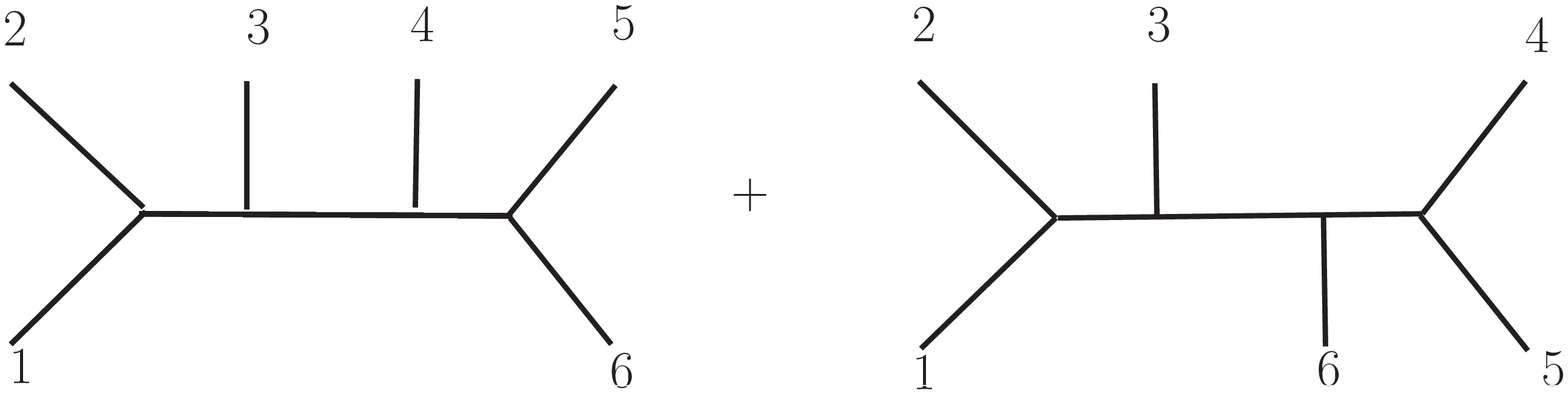}}
\centerline{\ninepoint \baselineskip=2pt {\bf Fig. 2.} {The diagrams associated
with $\langle {T_{123}\over s_1 t_1}\({V_4L_{56}\over s_5} + {L_{45}V_6\over s_4}\)\rangle$.}}
\smallskip
\bigskip
\centerline{\epsfxsize 3.4truein\epsfbox{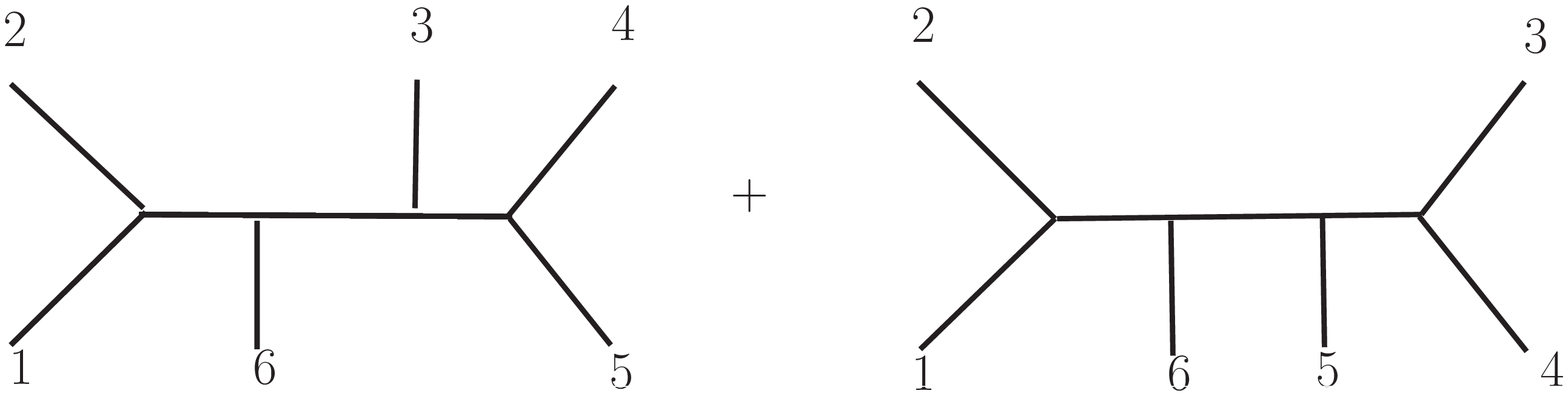}}
\centerline{\ninepoint \baselineskip=2pt {\bf Fig. 3.} {The diagrams associated
with $\langle {T_{126}\over s_1 t_3}\({V_3L_{45}\over s_4} + {L_{34}V_5\over s_3}\)\rangle$.}}
\smallskip

For the 7-point amplitude there are 6 diagrams which generate the 42 color-ordered cubic diagrams upon
cyclic symmetrization. The corresponding BRST-closed expression with the correct
pole structure is given by
\eqn\sevenpts{
{\cal A}_7(1,2,3,4,5,6,7) =
+ {\langle T_{231} L_{45}L_{67} \rangle \over s_2 t_1 s_4 s_6}
+ {\langle T_{123} T_{564} V_7  \rangle \over s_1 t_1 s_5 t_4}
+ {\langle T_{127} T_{345} V_6  \rangle \over s_1 t_7 s_3 t_3}
}
$$
- {\langle T_{123}T_{456}V_7 \rangle \over s_1 t_1 s_4t_4}
- {\langle T_{127}T_{453}V_6 \rangle\over s_1 t_7 s_4t_3}
- {\langle T_{123}L_{45} L_{67} \rangle \over s_1 t_1 s_4 s_6}
+ {\rm cyclic(1{\ldots} 7)}
$$
where $s_1,{\ldots} ,s_7$ and $t_1,{\ldots} ,t_7$ are the 7-point Mandelstam variables of
\stieS. The ten-dimensional 7-gluon expansion of \sevenpts\ contains more than 130 thousand terms \PSS\ 
and a few are written in appendix B.
As the results of \stieS\ are written in the four-dimensional helicity
formalism, a direct comparison with the results quoted there is not possible.

\vskip 15pt {\bf Acknowledgements:} 
I want to thank Nathan Berkovits, Oliver Schlotterer, Stephan Stieberger and Dimitrios Tsimpis
for discussions and for comments on the draft. I also thank the organizers of the Amsterdam String Theory Workshop 2010 for
the inspiring atmosphere, as the final details of this paper were worked out during those days. 
I thank the Werner-Heisenberg-Institut in M\"unchen for hospitality and partial financial
support, and in particular Stephan Stieberger and
Oliver Schlotterer for their invitations and warm hospitality.
I acknowledge support by the Deutsch-Israelische
Projektkooperation (DIP H52). 

\appendix{A}{The Bern-Carrasco-Johansson kinematic identities}

The 4-pt BCJ kinematic relation $n_u = n_s - n_t$ is mapped to the superspace
expression
$\langle L_{13}V^2V^4\rangle = \langle L_{12}V^3V^4\rangle - \langle L_{41}V^2V^3\rangle$.
Using $\langle L_{41}V^2V^3\rangle = -\langle L_{23}V^1V^4\rangle$ it can be rewritten as
\eqn\fptBCJ{
\langle  L_{\{12}V_{3\}} V^4\rangle = 0,
}
where $\{ijk\}$ means to sum over the cyclic permutation of the labels. Note that \fptBCJ\ 
can be explained from the fact that BRST-trivial quantities vanish. Explicitly,
\eqn\fptBCJexplained{
0= \langle Q(T_{123}V_4)\rangle = s \langle  L_{\{12}V_{3\}} V^4\rangle - (s+t+u)\langle L_{12}V_3V_4 \rangle,
}
which implies \fptBCJ\ because $s + t + u = 0$.

The 5-pt {\it extended BCJ relations} of \tye\vanhove\ are given by
\eqn\firstebcj{
  {n_4 - n_1 + n_{15}   \over s_{45}}
- {n_{10} - n_{11} + n_{13}   \over s_{24}}
- {n_3 - n_1 + n_{12} \over s_{12}}
- {n_5 - n_2 + n_{11}   \over s_{51}} = 0
}
\eqn\secebcj{
  {n_7 - n_6 + n_{14}   \over s_{14}}
- {n_{10} - n_{11} + n_{13}   \over s_{24}}
- {n_8 - n_6 + n_{9} \over s_{25}}
- {n_5 - n_2 + n_{11}   \over s_{51}} = 0
}
\eqn\threbcj{
  {n_{10} - n_9 + n_{15}   \over s_{13}}
+ {n_{5} - n_{2} + n_{11}   \over s_{51}}
- {n_4 - n_2 + n_{7} \over s_{23}}
+ {n_8 - n_6 + n_{9}   \over s_{25}} = 0
}
\eqn\fouebcj{
  {n_4 - n_1 + n_{15}   \over s_{45}}
- {n_{10} - n_{9} + n_{15}   \over s_{13}}
- {n_5 - n_2 + n_{11} \over s_{51}}
- {n_3 - n_5 + n_{8}   \over s_{34}} = 0.
}
Using the mappings of \mapping\ they become
\eqn\ebcju{
-{L_{45}\over s_{45}}L_{\{12}V_{3\}} 
+{L_{42}\over s_{24}}L_{\{13}V_{5\}}
-{L_{12}\over s_{12}}L_{\{34}V_{5\}}
+{L_{51}\over s_{51}}L_{\{23}V_{4\}} = 0,
}
\eqn\ebcjd{
-{L_{14}\over s_{14}}L_{\{23}V_{5\}} 
+{L_{42}\over s_{24}}L_{\{13}V_{5\}}
-{L_{25}\over s_{25}}L_{\{13}V_{4\}}
+{L_{51}\over s_{51}}L_{\{23}V_{4\}} = 0,
}
\eqn\ebcjt{
+{L_{13}\over s_{13}}L_{\{25}V_{4\}}
-{L_{51}\over s_{51}}L_{\{23}V_{4\}}
-{L_{23}\over s_{23}}L_{\{14}V_{5\}}
+{L_{25}\over s_{25}}L_{\{13}V_{4\}} = 0,
}
\eqn\ebcjq{
-{L_{45}\over s_{45}}L_{\{12}V_{3\}}
-{L_{13}\over s_{13}}L_{\{25}V_{4\}}
+{L_{51}\over s_{51}}L_{\{23}V_{4\}}
+{L_{34}\over s_{34}}L_{\{12}V_{5\}} = 0,
}
which one can check to hold true when expanding in components. Using
the momentum conservation relations
$$
s_{13} = s_{45} - s_{12} - s_{23}, \quad
s_{14} = s_{23} - s_{51} - s_{45}, \quad
s_{24} = s_{51} - s_{23} - s_{34}
$$
\eqn\momconser{
s_{25} = s_{34} - s_{12} - s_{51}, \quad
s_{35} = s_{12} - s_{45} - s_{34},
}
one finds that the LHS of \ebcju\ -- \ebcjq\ are BRST-closed.

\appendix{B}{The 5-, 6- and 7-gluon amplitudes}
The 5-gluon amplitude is easily obtained by using \PSS, and one can check
that the first few terms are
\eqn\fivegluon{
2880\, {\cal A}_5(1,2,3,4,5) =
}
       $$ - (k^1\cdot e^2)(k^1\cdot e^3)(k^1\cdot e^4)(e^1\cdot e^5)s_1^{-1}s_4^{-1} 
        + (k^1\cdot e^2)(k^1\cdot e^3)(k^1\cdot e^5)(e^1\cdot e^4)s_1^{-1} s_4^{-1} $$
       $$ - (k^1\cdot e^2)(k^1\cdot e^3)(k^2\cdot e^4)(e^1\cdot e^5)s_1^{-1} s_4^{-1} 
        + (k^1\cdot e^2)(k^1\cdot e^3)(k^2\cdot e^5)(e^1\cdot e^4)s_1^{-1} s_4^{-1} $$
       $$ - (k^1\cdot e^2)(k^1\cdot e^3)(k^3\cdot e^4)(e^1\cdot e^5)s_1^{-1} s_3^{-1} 
       + {\ldots} $$

The 6-gluon component expansion from the ansatz \sixptAnsatz\ generates
6706 terms of which the first few are \PSS
\eqn\sixgluon{
2880\,{\cal A}_6(1,2,3,4,5,6) = 
}
$$
       \big[ (k^1\cdot e^2)(k^1\cdot e^3)(k^1\cdot e^4)(k^1\cdot e^6)(e^1\cdot e^5)
       - (k^1\cdot e^2)(k^1\cdot e^3)(k^1\cdot e^4)(k^1\cdot e^5)(e^1\cdot e^6)
$$
$$
       - (k^1\cdot e^2)(k^1\cdot e^3)(k^1\cdot e^4)(k^2\cdot e^5)(e^1\cdot e^6)
       + (k^1\cdot e^2)(k^1\cdot e^3)(k^1\cdot e^4)(k^2\cdot e^6)(e^1\cdot e^5)
$$
$$       
       - (k^1\cdot e^2)(k^1\cdot e^3)(k^1\cdot e^4)(k^3\cdot e^5)(e^1\cdot e^6)        
       + (k^1\cdot e^2)(k^1\cdot e^3)(k^1\cdot e^4)(k^3\cdot e^6)(e^1\cdot e^5)\big] s_1^{-1}s_5^{-1}t_1^{-1}
$$
$$       
       - (k^1\cdot e^2)(k^1\cdot e^3)(k^1\cdot e^4)(k^4\cdot e^5)(e^1\cdot e^6) s_1^{-1}s_4^{-1}t_1^{-1} + {\ldots} 
$$
Similarly, the 7-gluon component expansion of \sevenpts\ has 134460 terms\foot{Some of those
terms contain $\e_{10}$ tensors and are expected to vanish once rules for the vanishing of
things like $\e_{10}^{[m_1{\ldots} m_{10}}\d^{m_{11}]}_n$ are implemented in \PSS.}
and the first ones
are
\eqn\setegluons{
2880\, {\cal A}_7(1,2,3,4,5,6,7) =
}
$$ \big[ + (k^1\cdot e^2)(k^1\cdot e^3)(k^1\cdot e^4)(k^1\cdot e^5)(k^1\cdot e^6)
     (e^1 \cdot e^7)
 - (k^1\cdot e^2)(k^1\cdot e^3)(k^1\cdot e^4)(k^1\cdot e^5)(k^1\cdot e^7)
     (e^1\cdot e^6) $$
       $$ + (k^1\cdot e^2)(k^1\cdot e^3)(k^1\cdot e^4)(k^1\cdot e^5)(k^2\cdot e^6)
       (e^1\cdot e^7)
  - (k^1\cdot e^2)(k^1\cdot e^3)(k^1\cdot e^4)(k^1\cdot e^5)(k^2\cdot e^7)
       (e^1\cdot e^6) $$
       $$ + (k^1\cdot e^2)(k^1\cdot e^3)(k^1\cdot e^4)(k^1\cdot e^5)(k^3\cdot e^6)
       (e^1\cdot e^7)\big] s_1^{-1}s_6^{-1}t_1^{-1}t_5^{-1} + {\ldots} $$
It is curious to note that the coefficient of $\pm 1/2880$ is the same for
all the terms in the 5-, 6- and 7-gluon amplitudes alike. This is the same 
coefficient which was observed in \treelevel\
to be the conversion factor required to match the RNS amplitudes at tree-level.

\appendix{C}{Shortcut to compute QL}

There is a shortcut to compute $QL$'s for $n$-points using only the $L$'s appearing
at $(n-1)$-points. The definitions of ${\tilde L}_{ij}$ and $L_{jiki}$ are \mafrabcj,
\eqn\Ldef{
V^i(z_i) U^j(z_j) \rightarrow { {\tilde L}_{ij}\over z_{ij}}, \quad\quad
{\tilde L}_{ij}(z_i)U^k(z_k) \rightarrow {L_{jiki}\over z_{ik}},
}
so that $Q{\tilde L}_{ij} = \lim_{z_j\rightarrow z_i}z_{ij}Q(V^i(z_i)U^j(z_j))$ and
$QL_{jiki} = \lim_{z_k\rightarrow z_i}z_{ik}Q({\tilde L}_{ij}(z_i)U^k(z_k))$ leads to
$$
Q{\tilde L}_{ij} = 
\lim_{z_j\rightarrow z_i} z_{ij}\p V^j(z_j)V^i(z_i) = - s_{ij}V^iV^j,
$$
$$
QL_{jiki} = -\lim_{z_k\rightarrow z_i}z_{ik}(s_{ij}V^i(z_i)V^j(z_i)U^k(z_k)
+ {\tilde L}_{ij}(z_i)\p V^k(z_k))
$$
\eqn\easy{
= -s_{ij}({\tilde L}_{ik}(z_i)V^j(z_i) + V^i(z_i){\tilde L}_{jk}(z_i))
+ (s_{ik}+s_{jk})V^k(z_i){\tilde L}_{ij}(z_i),
}
which agree with \trivial\ and \QL, respectively. In the above we used $QU^i(z) =
\p V^i(z) = \Pi^m(z)k^i_m V^i(z) + \p\theta^\a D_\a V^i(z) + \p\l^\a A^i_\a$,
which together with the OPE's of the conformal weight-one variables \refs{\siegel,\ictp} implies
that
\eqn\pVV{
\lim_{z_i\rightarrow z_j}Q(U^i(z_i)V^j(z_j)) = \lim_{z_i\rightarrow z_j}\p V^i(z_i) V^j(z_j) 
\rightarrow - s_{ij} {V^i(z_i) V^j(z_i)\over z_{ij}}.
}

\listrefs
\end